# Recommending Researchers in Machine Learning based on Author-Topic Model

Deepak Sharma[1], Bijendra Kumar[1], Satish Chand[2]

[1]Department of Computer Engineering,
Netaji Subash Institute of Technology, New Delhi, India
[2]School of Computer & Systems Sciences,
Jawaharlal Nehru University, New Delhi, India
{deepak.btg,bizender,schand20}@gmail.com

**Abstract.** The aim of this paper is to uncover the researchers in machine learning using author-topic model (ATM). We collect 16,855 scientific papers from six top journals in the field of machine learning published from 1997 to 2016 and analyze them using ATM. The dataset is broken down into 4 intervals to identify the top researchers and find similar researchers using their similarity score. The similarity score is calculated using Hellinger distance. The researchers are plotted using t-SNE, which reduces the dimensionality of the data while keeping the same distance between the points. The analysis of our study helps the upcoming researchers to find the top researchers in their area of interest.

## 1 Introduction

Topic models are a family of statistical models for discovering the latent topics in a group of documents [1]. In topic models, each topic is modeled as a probability distribution over words in the vocabulary of the corpus and each document in the corpus is modeled as a mixture of topics giving a multinomial distribution over the topics [2]. The probabilistic Latent Semantic Indexing (pLSI), proposed by Hofmann [3], works toward the probabilistic modeling of text; it however does not provide a probabilistic model at the level of documents. In order to overcome this problem, the latent Dirichlet allocation (LDA) model is discussed by Blei et al. [4], which is similar to the pLSI except the topic distribution in LDA is assumed to have a Dirichlet prior. The assumption usually brings about more reasonable mixtures of topics in a document. The LDA is a generative probabilistic model for collections of discrete data such as text corpora [5]. It is based upon the idea that the probability distribution over words in a document can be expressed as a mixture of topics, i.e., each document may be viewed as a mixture of various topics. LDA model can be viewed as a generative process. A document can be generated in following three steps as firstly to sample a mixture proportion from a Dirichlet distribution. Secondly, to sample a topic index according to the mixture proportion for each word in the document. Finally, to sample a word token from a multinomial distribution over the words specific to the sampled topic.

The document-topic and topic-word distributions learned by LDA describe the best topics for the documents and the most descriptive words for each topic. An extension of LDA is the author-topic model (ATM) [6, 7], which is based on the author-word model [8]. In ATM, a document is represented as a product of the mixture of topics of its authors, where each word is generated by the activation of one of the topics of an author of that document. ATM model has ability to give researchers a means to gain intuition about authorship and content in terms of topics at once. The authors associate with many kinds of metadata in documents, e.g. tags on posts on the web. This model can be used for author (or tag) prediction, data exploration, as features in machine learning pipelines, or to simply leverage the topic model with existing metadata. It provides a relatively simple probabilistic model for exploring the relationships between the authors, documents, topics, and words, wherein each author is represented by a multinomial distribution over topics and each topic is represented by a multinomial distribution over words. The words in a document co-authored by multiple authors are assumed to be the result of a mixture of topic mixture of each author. The topic-word and author-topic distributions are learned from text corpus. Compared to the LDA, the ATM provides the increase in salient topics and more reasonable researchers' interest patterns. It has indeed been proved to be an essential way to uncover the research interests of a researcher.

In this study, we uncover the researchers in machine learning research using the author-topic model and also find the similar researchers with their similarity score. This approach may be helpful to new researchers to find the most prominent researchers w.r.t. their research area for further research. The author of research papers are considered as researchers in our study. The rest of the paper is organized as follows. Section 2 provides an introduction of methodology adopted in our study and section 3 presents the result analysis. Finally, section 4 concludes the paper with some future direction.

## 2  An Introductory Discussion On Methods Adopted

In this section, we briefly discuss the Author-Topic model and process of finding similar researchers as these methods are used in our study.

### 2.1  Author-Topic Model (ATM)

The ATM is used to uncover the research interest of an author. The model is described by a set of linked probability distributions as follows:

$$\theta_a \sim Dir(\alpha) \tag{1}$$

$$\beta_k \sim Dir(\eta) \tag{2}$$

$$x_{dn} \sim Unif\left(\frac{1}{|A_d|}\right) \tag{3}$$

$$z_{dn} \sim Mult(\theta_a, x_{dn} = a) \quad (4)$$

$$w_{dn} \sim Mult(\beta_k, z_{dn} = k) \quad (5)$$

where, Eq. (3). says that the author of a word $w_{dn}$ is drawn uniformly with probability one over the number of authors in document $d$. Eq. (4). says that we draw $z_{dn}$ from $\theta_a$, assuming $x_{dn} = a$. The intuition behind each of these parameters are discussed below:

- $\theta_a$ is a probability vector such that $\theta_{ak}$ is the probability that author $a$ writes about topic $k$.
- $\beta_k$ is a probability vector such that the probability that word $v$ is used in topic $k$ is equal to $\beta_{kv}$.
- $x_{dn}$ is a latent variable to indicate which author is responsible for word $n$ in document $d$.
- $z_{dn}$ is also a latent variable to indicate which topic generates word $n$ in document $d$.

Figure 1 shows the graphical model of ATM. We can interpret the edges in graph as dependency between two variables, e.g., $z_{dn}$ depends on $\theta_a$, and the absence of an edge represents conditional independence, e.g., when conditioned on $z_{dn}$, $w_{dn}$ is independent of $\theta_a$, i.e. $p(w_{dn}|\beta_k, z_{dn} = k)$ does not depend on $\theta_a$.

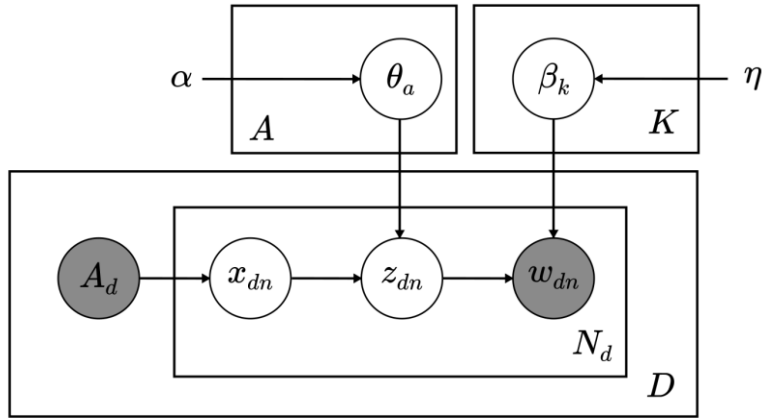

**Fig. 1.** Graphical model of author-topic model

The author-topic model may be viewed as a process that generates words in a document based on the authors of that document.
- For each author $a \in \{1, ..., A\}$, construct Eq. (1).
- For each topic $k \in \{1, ..., K\}$, construct (2).
- For each document $d \in \{1, ..., D\}$:

- Given document $d's$ authors, $A_d$.
- For each word in document $n \in \{1, \ldots, N_d\}$.
  - Assign an author to current word by constructing Eq. (3).
  - Conditioned on $x_{dn}$, assign a topic by constructing Eq. (4).
  - Conditioned on $z_{dn}$, assign a topic by constructing Eq. (5).

The Variational Bayes (VB) method is used for approximate inference in the author-topic model to maximize the lower bound on the log likelihood. Though the lower bound and the conditional likelihood tell how well the algorithm converges, yet they don't tell much about the quality of the topics [9]. Mimno [10] discusses the domain experts annotated quality of topics from a trained LDA model. They devise a measure to be strongly correlated with the human annotator's judgment of the topic quality, which is the topic coherence.

### 2.2 Finding Similar Researchers

Here, we set up a system that takes the name of researchers as input and produces the researchers that are most similar, which is used as a component in information retrieval (i.e. a search engine of finding similar researchers). The common similarity measure between two researchers is computed using the cosine distance. We however use the Hellinger distance as it is a natural way of measuring the distance (i.e. dissimilarity) between two probability distributions, which is defined in discrete version as follows:

$$H(a_1, a_2) = \frac{1}{\sqrt{2}} \sqrt{\sum_{i=1}^{K} (\sqrt{a_{1i}} - \sqrt{a_{2i}})^2} \qquad (6)$$

where $a_1$, $a_2$ are topic distributions for two different researchers, $K$ is number of topics in author-topic model.
We define the similarity between authors $a_1$ and $a_2$ as follows:

$$S(a_1, a_2) = \frac{1}{1 + H(a_1, a_2)} \qquad (7)$$

## 3 Analysis of Our Result

Here, we first look at the dataset used for our study and preprocess it before applying the Author-topic model and then plot the researchers to explore the author-topic representation. Finally, we present the researchers with highest topic distribution with their similar researchers in machine learning research.

### 3.1 Dataset Description

We start collecting the data by preparing a list of appropriate well-known journals that publish the high-quality research in machine learning. Figure 2 shows the distribution of number of research papers included in our study in different colors as shown in legends.

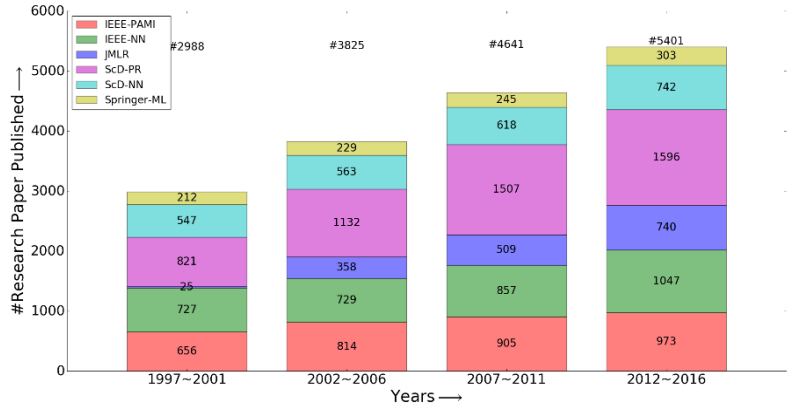

**Fig. 2.** Distribution of number of research papers in our study

We have included well-established journals like Springer machine learning (Sp-ML), ScienceDirect neural networks (ScD-NN), ScienceDirect pattern recognition (ScD-PR), Journal of machine learning research (JMLR), IEEE transactions on neural networks (IEEE-NN) and IEEE transactions on pattern analysis and machine intelligence (IEEE-PAMI). The dataset is broken into 4 intervals such as 1997~2001, 2002~2006, 2007~2011. 2012~2016. The titles, authors, and abstracts of the research papers have been taken from the electronic library of the above-mentioned journal articles. We consider only journal articles in our study.

### 3.2  Preprocessing Data

The preprocessing of the research papers is done by tokenizing the text including *title* and *abstract*, replace all whitespace by single spaces, remove all punctuation and numbers, filtering terms (words), by applying a custom stop word list along with the original stop words of the SMART system [11], which is a common preprocessing step in text mining. Then apply stemming [12] to words, to perform their lemmatization, add multi-word named entities, add frequent bigrams and remove frequent and rare words. Construct a mapping from author names to documents as required by ATM. Finally, the vectorized representation of the text by computing bag-of-words with authors mapping is given to ATM.

### 3.3  Training the Author-topic model

We train the author-topic model on the preprocessed data prepared in the above sections. For our experimental purpose, we use the Gensim as a pure Python library for implementing the ATM, which is an NLP software framework created on the idea of document streaming [13]. This framework has two objectives, first one represents the indexing of digital document and similarity search, and second one represents the memory-efficient, fast and scalable algorithms for Singular Value Decomposition for unsupervised learning and semantic analysis of plain text in digital collections. It requires the open source NumPy for n-dim array manipulation and SciPy for numerical integration and optimization. The advantages of Gensim are fast processing of large datasets and memory independence because the term-by-document matrix does not have to be stored in memory. In our experiments of ATM, the number of topics is fixed as 5 for each dataset, the symmetric Dirichlet priors $\alpha$ and $\beta$ are set as 0.5 and 0.1, respectively, and VB is run for 2000 iterations. We have run the model for same setting with different random states for five times and chosen the best model for further analysis. Table 1 shows the statistical details of our best ATM.

**Table 1.** Statistical details our best ATM

| Dataset | #Research Papers | #Unique Tokens | #Authors | Topic Coherence | Word per-bound |
|---|---|---|---|---|---|
| 1997~2001 | 2988 | 1862 | 5368 | -1.572e+03 | -6.653556 |
| 2002~2006 | 3825 | 2219 | 7236 | -1.657e+03 | -6.766959 |
| 2007~2011 | 4641 | 2704 | 9542 | -1.576e+03 | -6.871249 |
| 2012~2016 | 5401 | 3035 | 11761 | -1.684e+03 | -6.713045 |

### 3.4 Plotting the Researchers

Here we explore the author-topic representation in an intuitive manner. We take all the author-topic distributions and embed them in a 2-D space by reducing the dimensionality of this data using the t-SNE. The t-SNE is a method that attempts to reduce the dimensionality of a dataset, while maintaining the distances between the points. If two researchers are close together in plot below, then their topic distributions would be similar. Figures 3(a)-3(d) represent plots of researchers for dataset in different time-period in which the circles refer to individual researchers and their sizes represent the number of documents attributed to the corresponding author. Large clusters of researchers tend to reflect some overlap in interest. As evident from these figures, the number of researchers has increased in machine learning in successive years, indicating the popularity of this area.

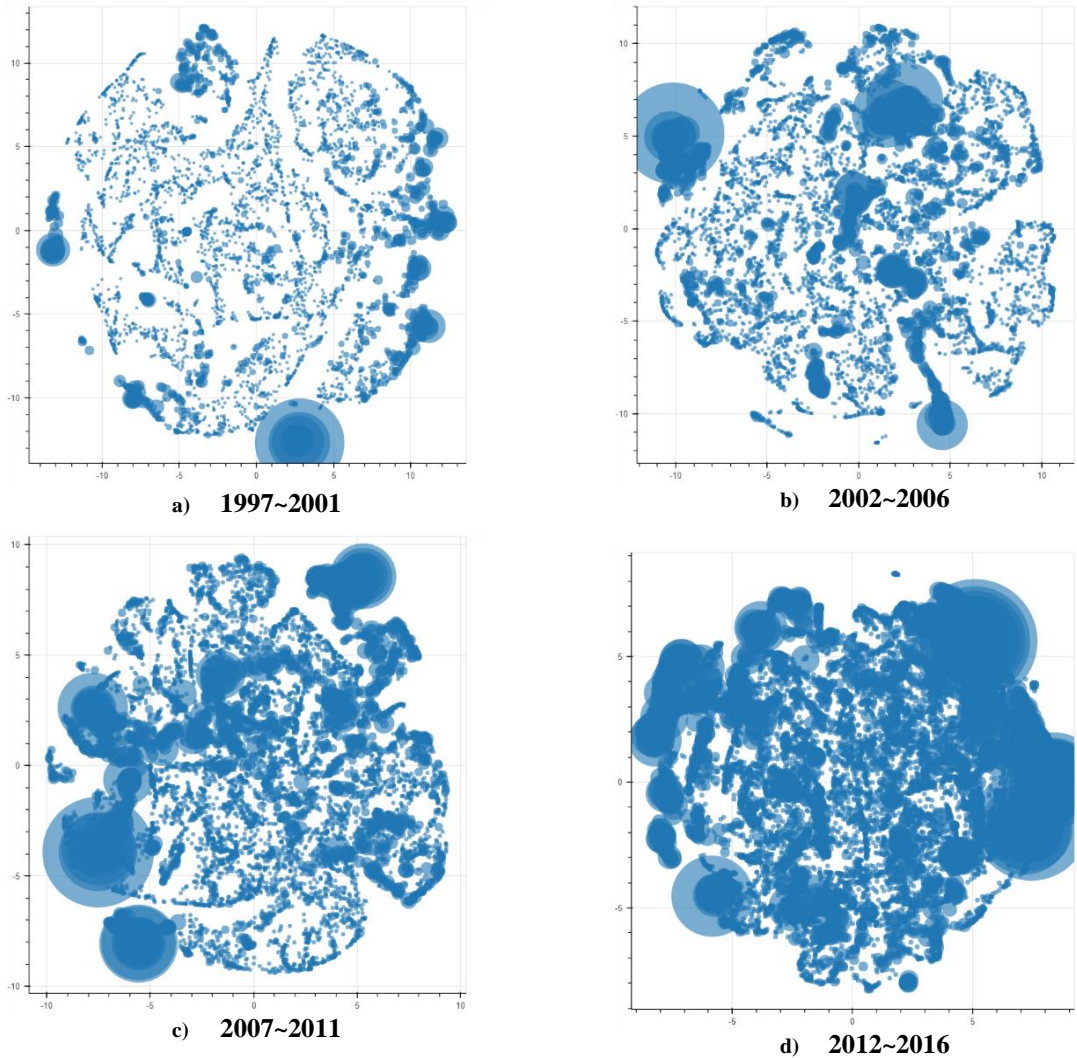

**Fig. 3.** Plotting of Authors for a) 1997~2001, b) 2002~2006,
c) 2007~2011, and d) 2012~2016

### 3.5   Identifying Researchers in Machine Learning

The researchers for each topic with highest topic distribution, along with their similar researchers of same interest are discussed in this section. Table 2 represents the list of the researchers with their similar researchers of each topic for the time period 1997~2001. Each researcher represents his topic distribution and top 10 words in each topic. The top researcher is selected based on the highest topic distribution.

Then, the similar researchers are evaluated on the similarity score based on Hellinger distance as discussed in section 2.2.

**Table 2.** List of the researcher with their similar researchers of each topic for the period 1997~2001.

| Topic Id | Top 10 words in Topic | Top Researchers Researcher Name | Topic % | Top Five Similar Researchers Researcher Names (Similarity Score) |
|---|---|---|---|---|
| T1 | imag, recognit, featur, segment, propos, match, base, detect, transform, shape | Anil K. Jain | 0.9991 | Hong Yan (0.999876), Kuo-Chin Fan (0.999498), Venu Govindaraju (0.999328), B. B. Chaudhuri (0.998879), Dinggang Shen (0.998863) |
| T2 | model, object, motion, surface, imag, visual, 3d, estim, comput, view | Stephen Grossberg | 0.9972 | Gregory D. Hager (0.998596), Dmitry Goldgof (0.997027), Larry S. Davis (0.995845), Dimitris Metaxas (0.995644), Joseph Weber (0.995015) |
| T3 | neural, network, function, control, neural_network, system, model, propos, neuron, nonlinear | Jun Wang | 0.9980 | Tianping Chen (0.999380), Youshen Xia (0.998871), Kazuyuki Aihara (0.998310), S. Matsuda (0.997719), JL Diaz De Leon (0.997432) |
| T4 | learn, problem, fuzzi, model, train, predict, general, rule, perform, number | Roni Khardon | 0.9955 | Manfred K. Warmuth (0.999729), Shai Ben-David (0.999643), G. Bebis (0.997133), Luc De Raedt (0.996080), Yu-Tsung Wang (0.995384), |
| T5 | data, classif, set, classif, function, estim, model, cluster, pattern, problem | Robert P. W. Duin | 0.9957 | V. Tresp (0.997860), Nathan S. Netanyahu (0.997814), Ludmila I. Kuncheva (0.997117), A. D. Back (0.996863), S. Chen (0.995902) |

The topic label of five topics in Table 2 are image processing, video processing, neural network, fuzzy model, and supervised/unsupervised learning. It also contains the top researchers selected based on its highest topic proportion among researchers and also the top five similar researchers, selected for each top researcher.

Table 3 represents the list of researchers along with their similar researchers of each topic for the time period 2002~2006. The topic label of five topics in Table 3 are image processing, neural network, supervised learning, face recognition and unsupervised learning. The researcher, Jung Wang was the lead researcher in neural network and amazingly the researcher Stephen Grossberg was the lead researcher in video processing at time period 1997~2001, who has become the most similar researcher in the neural network during the time period 2002~2006.

Table 3. List of the researcher with their similar researchers of each topic for the period 2002~2006.

| Topic Id | Top 10 words in Topic | Top Researchers Researcher Name | Topic % | Top Five Similar Researchers Researcher Names (Similarity Score) |
|---|---|---|---|---|
| T1 | imag, feature, recognit, system, extract, perform, detect, comput, motion, estim | P. Raveendran | 0.9914 | Ehud Rivlin (0.999776), Guang Jiang (0.998585), Hung-Tat Tsui (0.997652), Jan Flusser (0.997599), Hongjiang Zhang (0.996974) |
| T2 | neural, model, network, system, control, learn, function, neuron, dynam, neural_network | Jun Wang | 0.9978 | Stephen Grossberg (0.999891), S. S. Ge (0.999524), Xing-Bao Gao (0.999319), Masato Okada (0.999242), T. Yagi (0.998953) |
| T3 | model, learn, problem, train, estim, error, set, data, classif, perform | Chih-Jen Lin | 0.9973 | Bernhard Schölkopf (0.999322), Edward Dougherty (0.999211), Ashwin Srinivasan (0.999130), Massimiliano Pontil (0.998529), Sumio Watanabe (0.998463) |
| T4 | imag, object, model, segment, shape, match, face, region, point, color | Katsushi Ikeuchi | 0.9970 | Kenneth Kin-Man Lam (0.999233), Isabelle Bloch (0.999175), F. Sandoval (0.996433) N. Kiryati (0.996215), Jian Kang Wu (0.993983) |
| T5 | data, cluster, featur, analysi, classif, set, vector, space, classify, linear | Haesun Park | 0.9958 | Songcan Chen (0.999993), Jian Yang (0.999441), Sung-Yang Bang (0.998959), Marco Loog (0.998843), Peg Howland (0.998770) |

Table 4 represents the list of the researchers along with their similar researchers of each topic for the time period 2007~2011. The topic labels of five topics in Table 4 are optimization/approximation for neural network, linear/non-linear classification, face recognition, deep neural network and image/video processing. During this time period, new researchers have made their presence. During this time, the machine learning research has evolved and more optimization of the neural network as deep learning was presented in the research community.

**Table 4.** List of the researcher with their similar researchers of each topic for the period 2007~2011.

| Topic Id | Top 10 words in Topic | Top Researchers Researcher Name | Topic % | Top Five Similar Researcher Researcher Names (Similarity Score) |
|---|---|---|---|---|
| T1 | function, problem, optim, point, network, neural, approxim, linear, solut, comput | Z. Wang | 0.9985 | Jinde Cao (0.999189), W. X. Zheng (0.999016), Joviša Zunić (0.998782), P. Shi (0.998208), Q. L. Han (0.998118) |
| T2 | learn, data, classif, classifi, problem, set, kernel, perform, train, model | Johan A. K. Suykens | 0.9971 | Chih-Jen Lin (0.999605), Klaus-Robert Müller (0.999405), X. Yao (0.999284), Alberto Suárez (0.999209), Xiaotong Shen (0.999134) |
| T3 | feature, imag, recognit, cluster, data, face, perform, extract, local, analysi | David Zhang | 0.9988 | Lei Zhang (0.999794), Daoqiang Zhang (0.999335), Dacheng Tao (0.999248), Bir Bhanu (0.999150), Thomas S. Huang (0.998996) |
| T4 | system, model, neural, dynam, adapt, control, process, neuron, learn, perform | Ganesh Kumar Venayagamoorthy | 0.9936 | Justin C. Sanchez (0.999579), T. H. Lee (0.998875), Keiichiro Inagaki (0.998724), R. Wang (0.998409), K. C. Tan (0.997953) |
| T5 | imag, model, object, segment, detect, approach, estim, graph, motion, structur | Long Quan | 0.9985 | Michael S. Brown (0.999954), Rene Vidal (0.999715), Stephen Lin (0.999715), Ruigang Yang (0.999488), Ronald Chung (0.999341) |

Table 5 represents the list of the researchers along with their similar researchers of each topic for the time periods 2012~2016. The topic labels of five topics in Table 5 are computer vision, support vector machine, optimizing the neural network, face recognition and unsupervised learning. The researcher Jinde Cao was the lead researcher in neural network and amazingly we observe on the researcher that Jun Wang as the most prominent researcher in neural networks during the time periods 1997~2001, 2002~2006 and 2012~2016. Similarly, the researcher Isabelle Bloch was one of the researcher in face recognition and has become the researcher in computer vision during the time period 2012~2016.

**Table 5.** List of the researcher with their similar researchers of each topic for the period 2012~2016.

| Topic Id | Top 10 words in Topic | Top Researchers Researcher Name | Topic % | Top Five Similar Researchers Researcher Names (Similarity Score) |
|---|---|---|---|---|
| T1 | imag, detect, object, segment, recognit, feature, system, visual, approach, human | Ashutosh Saxena | 0.9939 | Isabelle Bloch (0.999845), Kaiming He (0.999821), Jun Zhang (0.999694), Guodong Guo (0.999331), H. Tang (0.999294) |
| T2 | comput, vector, effici, support, approach, machin, train, kernel, time, optim | Yuan-Hai Shao | 0.9940 | Da-Han Wang (0.999561), Zhuang Wang (0.997633), Irfan Ahmad (0.997278), Yali (0.995624), Jialei Wang (0.995135) |
| T3 | neural, network, function, control, system, estim, optim, learn, condit, bound | Jinde Cao | 0.9988 | Jun Wang (0.999828), Tingwen Huang (0.999745), S. Jagannathan (0.999697), H. Zhang (0.999656), W. X. Zheng (0.999542) |
| T4 | imag, face, local, feature, recognit, graph, robust, estim, spars, shape | David Zhang | 0.9974 | J. Yang (0.999626), Mingbo Zhao (0.999444), Y. Xu (0.998836), In So Kweon (0.998639), Y. Xiang (0.998565) |
| T5 | learn, classif, cluster, feature, set, classifi, select, label, class, approach | R. Polikar | 0.9981 | G. Rosen (0.999951), Licheng Jiao (0.999759), Zhi-Hua Zhou (0.999277), Steven C. H. Hoi (0.999210), Jun Zhu (0.999178), |

In this section, we have presented the results of our study and the top researchers along with similar researchers during different time periods. This study has facilitated new researchers to find the prominent researchers in their field of interest.

## 4  Conclusions and Future Work

In this study, we have performed the identification of researchers in machine learning for last two decades depending on the research articles published in six well-known journals. In order to understand the researchers' interest on topics for 20 years, we divided the dataset into four sets for the time periods of 1997~2001, 2002~2006, 2007~2011, 2012~2016 and performed the analysis using ATM to select 5 topics and related words corresponding to the topics and finding the top 5 similar researchers for top researcher on each topic. This study can be helpful the new researchers in machine learning area to get insight of top researchers related to their area of interest. In future, we intend to extend this study to find the research communities depending on the research interest and analyze the research trends in communities in different time periods.